\pdfoutput=1
\documentclass[conference]{IEEEtran}

\def\BibTeX{{\rm B\kern-.05em{\sc i\kern-.025em b}\kern-.08em
    T\kern-.1667em\lower.7ex\hbox{E}\kern-.125emX}}

\usepackage{booktabs}  
\usepackage{pbox}
\usepackage{epsfig,endnotes}
\usepackage{epstopdf}
\usepackage{xcolor}
\usepackage{graphicx}
\usepackage{xspace}
\usepackage{enumitem}
\usepackage{url}
\usepackage{balance}
\usepackage[utf8]{inputenc}
\usepackage{setspace}
\usepackage[tableposition=top,font={small}, skip=7pt]{caption}  
\usepackage{arydshln}
\usepackage{amsmath,amssymb,amsfonts}
\usepackage{algorithmic}
\usepackage{textcomp}

\newcommand{\point}[1]{\vspace{0.01in}\par\vspace{0.01in}\noindent{\textbf{#1:} }}
\newcommand{\toolname}{FNDaaS\xspace}
\newcommand{\eg}{{e.g.,}\xspace}
\newcommand{\etc}{{etc.}\xspace}

\newcommand{\ie}{{i.e.,}\xspace}

\newcommand{\realNewsSet}{1183\xspace}
\newcommand{\fakeNewsSet}{637\xspace}
\newcommand{\totalSites}{1820\xspace}
\newcommand{\extension}{\toolname-client\xspace}
\newcommand{\numOfFeatures}{187\xspace}

\begin{document}

\title{\toolname: Content-agnostic Detection of Websites Distributing Fake News}

\author{\IEEEauthorblockN{Panagiotis Papadopoulos}
\IEEEauthorblockA{\textit{Telefonica Research}
\\Spain} 
\and
\IEEEauthorblockN{Dimitris Spythouris}
\IEEEauthorblockA{\textit{University of Crete}
\\Greece} 
\and
\IEEEauthorblockN{Evangelos P. Markatos}
\IEEEauthorblockA{\textit{University of Crete \& FORTH-ICS}
\\Greece} 
\and
\IEEEauthorblockN{Nicolas Kourtellis}
\IEEEauthorblockA{\textit{Telefonica Research}
\\Spain} 
}


\maketitle

\begin{abstract}
    Automatic fake news detection is a challenging problem in misinformation spreading, and it has tremendous real-world political and social impacts.
    Past studies have proposed machine learning-based methods for detecting such fake news, focusing on different properties of the published news articles, such as linguistic characteristics of the actual content, which however have 
limitations due to the apparent language barriers. 
    Departing from such efforts, we propose \emph{Fake News Detection-as-a-Service} (\toolname), the first automatic, content-agnostic fake news detection method, that considers new and unstudied features such as network and structural characteristics per news website.
    This method can be enforced \textit{as-a-Service}, either at the ISP-side for easier scalability and maintenance, or user-side for better end-user privacy.
    We demonstrate the efficacy of our method using more than 340K datapoints crawled from existing lists of \fakeNewsSet fake and \realNewsSet real news websites, and by building and testing a proof of concept system that materializes our proposal.
    Our analysis of data collected from these websites shows that the vast majority of fake news domains are very young and appear to have lower time periods of an IP associated with their domain than real news ones.
    By conducting various experiments with machine learning classifiers, we demonstrate that \toolname can achieve an AUC score of up to 0.967 on past sites, and up to 77-92\% accuracy on newly-flagged ones.
\end{abstract}

\section{Introduction}
\label{sec:introduction}

The rise of fake news, and the proliferation of doctored narratives spread by humans and/or bots online, are challenging web publishers, but also the society as a whole.
Misinformation  not only poses serious threats to the integrity of journalism, but has also created societal turmoils (i) in the economy: \eg the maliciously reported death of Ethereum's founder Vitalik Buterin caused a market value loss of \$4 billion~\cite{ethereumDeath},
and (ii) in the political world: \eg the influence of Russia in the 2016 American elections through fast-spreading fake news~\cite{lakhta}, the Brexit vote in the United Kingdom~\cite{10.1177/0894439317734157} and the tumultuous US presidential election highlighted how the digital age has affected news~\cite{bovet2019influence, 10.1080/10999922.2017.1285540}.
But fake news may even have an impact on human life too: \eg PizzaGate during the 2016 US presidential elections
~\cite{pizzagate}, the unfounded concerns about a link between the MMR vaccine and autism 
~\cite{Ramsay752}), the deaths~\cite{islam2020covid} caused by
COVID-19-related misinformation.

Evidently, the undoubted impact of fake news in our well-being has raised a lot of concerns among the community.
BBC interviewed a panel of 50 experts: propaganda and fake news~\cite{bbc} were among the ``grand challenges we face in the 21st century''.
Additionally, a US study~\cite{pew} found that 64\% of adults believe fake news stories cause a great deal of confusion, and 23\% said they had shared fabricated political stories themselves, by mistake or intentionally.

This grand challenge has also led to an abundance of initiatives from journalists, moderators and experts (\eg DisinformationIndex~\cite{DisinformationIndex}, Snopes~\cite{snopes}, MBFC~\cite{mbfc}, Checkkology~\cite{Checkology}, Factcheck.org~\cite{factcheck}, CrossCheck~\cite{CrossCheck}), or academics ~\cite{10.14778/3137765.3137815, 10.1145/3308560.3316460,9101457,10.1145/3437963.3441828} aiming to (manually) curate, verify and fact check content and claims of news sites.
However, considering the sheer volume of fake news sites spawning in a short period of time, and especially during important events (\eg the USA presidential elections~\cite{10.1145/3447535.3462510}, the COVID-19 pandemic~\cite{covidMisinf} or the war in Ukraine~\cite{warukraine}), it is apparent that \emph{manual fact checking by journalists and other experts is not scalable}.

Interestingly, there is a significant body of research aiming to provide automated misinformation detection by exploiting textual or content characteristics of published articles~\cite{9101457, ott-etal-2013-negative, mahyoob2020linguistic, oraby-etal-2015-thats, singh2017automated, 10.5555/2390665.2390708}. 
Albeit important proposals, such approaches based on linguistic analysis have specific limitations arising from the language barriers; 
automated machine-learning-based fake news detection systems need to be trained on text in a specific language to be highly performing.
However, with the majority of annotated articles and other computational resources in data and models focused on the English language, a large portion of news websites across the world are left unchecked.

In this work, we propose \emph{Fake News Detection-as-a-Service} (\toolname): the first of its kind, holistic system designed to detect websites disseminating fake news, that is completely content-agnostic.
\toolname does not take into account the actual fake news text (content) of articles shared on these sites; instead, it relies on the network and structural characteristics of the website sharing them (\eg DNS record changes, domain age, domain expiration and re-registration patterns, DOM characteristics, number of HTTP redirections, page rendering timings, \etc).
The goal of \toolname is not only to reduce the effort of manual curators (by providing a short list of suspicious websites), but eventually to provide an autonomous service that parses the Web and produces frequent reports (\ie lists) of websites classified as potentially sharing fake news, that browsers can utilize locally (via a browser extension) and warn users accordingly.
The contributions of this paper are summarized as follows:
\begin{enumerate}
    \item We design \toolname: a novel holistic fake news detection service that breaks the linguistic barriers and detects websites potentially sharing fake news, without relying on the website's content, but instead leveraging the sites' network and structural characteristics. 

    \item We evaluate the feasibility and effectiveness of our approach by using \numOfFeatures features extracted from a set of \fakeNewsSet fake and \realNewsSet real news websites (total of 340K datapoints). We compare websites sharing fake and real news in features important for the machine learning (ML) classification, and identify key differences with respect to network and structural properties of these websites. Our ML performance results show that \toolname achieves high Precision, Recall, and F1 scores (all above 91\%) and an AUC score of 0.967, using the top 35 out of a total of \numOfFeatures extracted features. Our approach is also applicable in a real-time setting, where 27 features available only at run-time are used, with 82\% Precision and Recall and 0.86 AUC.
    We also apply these classifiers trained on older data to websites newly-flagged as serving real or fake news, with 77-92\% accuracy.
    
    \item We implemented a prototype of \toolname as a cloud service that automatically crawls and classifies websites on the web. The results of this classification are transmitted periodically to a browser extension on the user's end that is responsible for enforcing and warning the users about the visited website.  We make our prototype available for further experimentation of this research\footnote{Link to the open source repositories for the: (1) server  \url{https://github.com/dimspith/fn-api} and (2) browser extension \url{https://github.com/dimspith/fn-blacklist}}.

    \item Early system evaluation of our prototype shows that the client component has minimal effect on the user experience (consuming less than 50 MBytes of memory), and takes less than 400 ms to warn the user about the marked-as-fake news website they are attempting to visit.
\end{enumerate}
\section{Related Work}
\label{sec:related}

In~\cite{conroy2015automatic}, authors provide a typology of varieties of veracity assessment methods emerging from: (i) linguistic cue (ML-based) approaches, and (ii) social network analysis approaches. 
In~\cite{depaulo2003cues}, authors explore linguistic features by analyzing cues of fake stories from a physiological point of view and find that fake stories tend to contain an unusual language. In~\cite{papadogiannakis2022leveraging} and~\cite{papadogiannakis2022funds}, authors use a graph clustering approach to analyze administration and monetization mechanisms of fake news sites that operate in groups.
In~\cite{10.1145/2872427.2883085}, authors analyzed Wikipedia hoaxes to measure how long they survive before being debunked, how many page-views they receive, and how heavily they are referred to by documents on the Web. 
In~\cite{10.5555/2002472.2002512}, authors perform deception detection by employing Linguistic Inquiry and Word Count (LIWC) to investigate the role of individual words in a document.
In~\cite{karimi2018multi}, authors set out to explore (a) how to effectively combine information from multiple sources for fake news detection and (b) how to mathematically discriminate between degrees of fakeness.
In~\cite{10.1145/3132847.3132877}, authors propose a model that achieves an accurate and automated prediction by combining: (a) its text, (b) the user response it receives, and (c) the source users promoting it.
FACE-KEG in~\cite{10.1145/3437963.3441828} aims to provide explainable fact checking. It constructs a relevant knowledge graph for a given input fact or claim using a large-scale structured knowledge base, and it also retrieves and encodes relevant textual context about the input text from the knowledge base.

In~\cite{shu2019context}, authors argue about the role of social context for fake news detection and propose a tri-relationship that models publisher-news relations and user-news interactions simultaneously for fake news classification.
In\cite{10.1145/3459637.3482139}, authors (i) design a benchmark of fake news dataset for multi-domain fake news detection with domain label annotated, and  (ii) propose an effective multi-domain Fake News detection model by utilizing domain gate to aggregate multiple representations extracted by a mixture of experts.
In \cite{10.1145/3442381.3450111}, authors examine the performance of a broad set of modern transformer-based language models and show that with basic fine-tuning, these models are competitive with and can even significantly outperform state-of-the-art methods. 

Check-it~\cite{paschalides2019check} constitutes an ensemble method that combines different signals (domain name, linguistic features, reputation score and others) to generate a flag about a news article or social media post.
The NELA toolkit~\cite{horne2018assessing} enables users to explore the credibility of news articles by using well-studied content-based markers of reliability and bias, as well as, filter and sort through article predictions.
In~\cite{10.1145/3447535.3462510}, authors conduct a temporal and traffic analysis of the network characteristics of fake new sites (user engagement, lifetime, \etc) and propose a methodology to study how websites may be synchronizing their alive periods, and even serving the exact same content for months at a time.
In~\cite{8919302}, authors propose a methodology to detect fake news by exploiting both the textual and visual features of a given article.

In~\cite{perez2017automatic}, authors introduce two datasets covering seven different news domains.
They (i) describe the collection, annotation, and validation process in detail and present several exploratory analyses on the identification of linguistic differences in fake and legitimate news content, and (ii) conduct a set of learning experiments to build accurate fake news detectors.
FakeDetector~\cite{9101457} is an automatic fake news credibility inference model for social networks that: (i) extracts a set of explicit and latent features from the  textual information, (ii) builds a deep diffusive network model to learn the representations of news articles, creators and subjects simultaneously.
Similar to our work, in their preliminary study~\cite{10.1145/3308560.3316460}, authors  characterize the bias of news websites, \ie identify portals with the same political orientation, using a link-based approach to show that a community detection algorithm can identify groups formed by sources with the same political orientation.
Finally, in~\cite{Bozarth_Budak_2020}, authors highlight the need to move toward systematic benchmarking, considering the lack of a comprehensive model evaluation framework that can provide multifaceted comparisons between fake news classification models beyond the simple evaluation metrics (\eg accuracy or f1 scores). 

\point{Comparison} Our proposed \toolname method is novel as it considers new and unstudied features per website.
In particular, it does not rely on the content, reputation, authors, sentiment, \etc, of the article, as previously mentioned studies do.
Instead, our content-agnostic method relies on the website's network and structural characteristics that can be readily available while the website is being browsed by the end-user.
Furthermore, our content-agnostic approach is not limited by language barriers.
Finally, we outline how the \toolname method can be applied in two ways: server-side or user-side, with different advantages in each case, but primarily high scalability across many users and lower maintenance effort in the server-side, and higher privacy in the user-side.

\section{\toolname System Design}
\label{sec:design}

In this section, we present the dual design of our approach which (based on how the filtering of fake news is enforced) can be deployed either (i) at the user side, or (ii) at the ISP side.
Of course, the key component in both methods is the fake news classifier module which includes data collection and preprocessing, as well as the machine learning modeling parts.

\begin{figure}[t]
    \centering
    \begin{minipage}[t]{0.49\textwidth}
    \centering
    \includegraphics[width=0.6\textwidth]{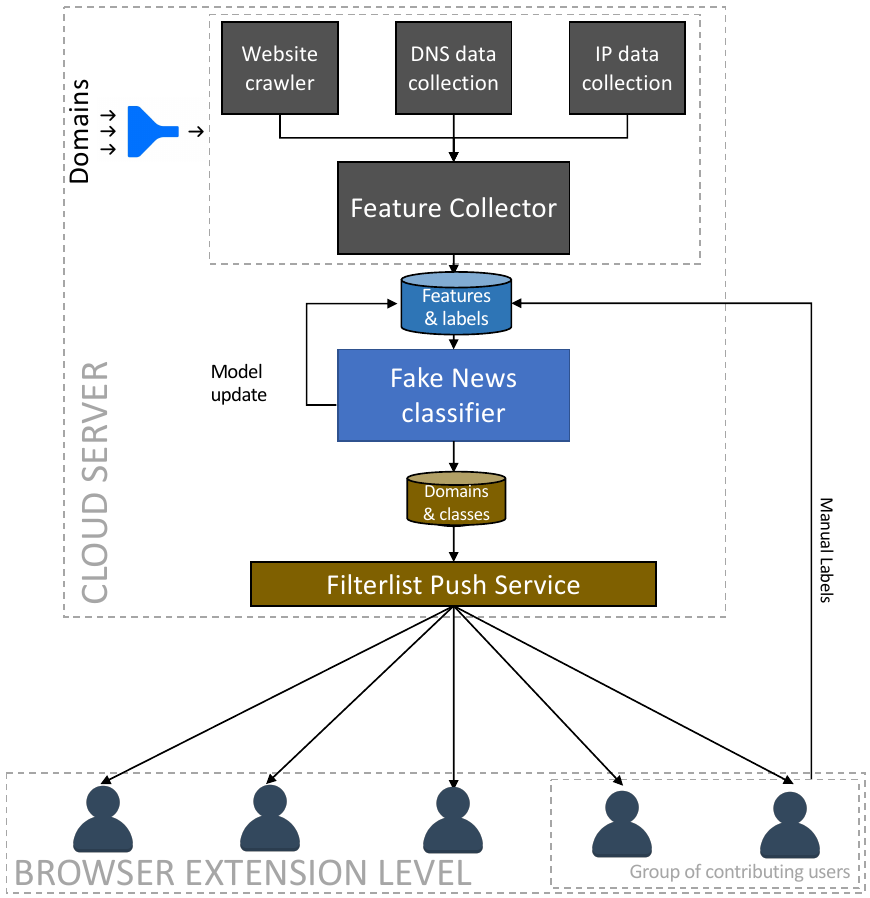}
    \caption{High level overview of \toolname when deployed with filtering performed at user side via browser plugin.
    Manual labels of websites can also be contributed by specialized users for continuous re-training and tuning of the ML model and filterlist.}
    \label{fig:design1}
    \end{minipage}
    \hfill
    \begin{minipage}[t]{0.49\textwidth}
    \centering
    \includegraphics[width=0.9\textwidth]{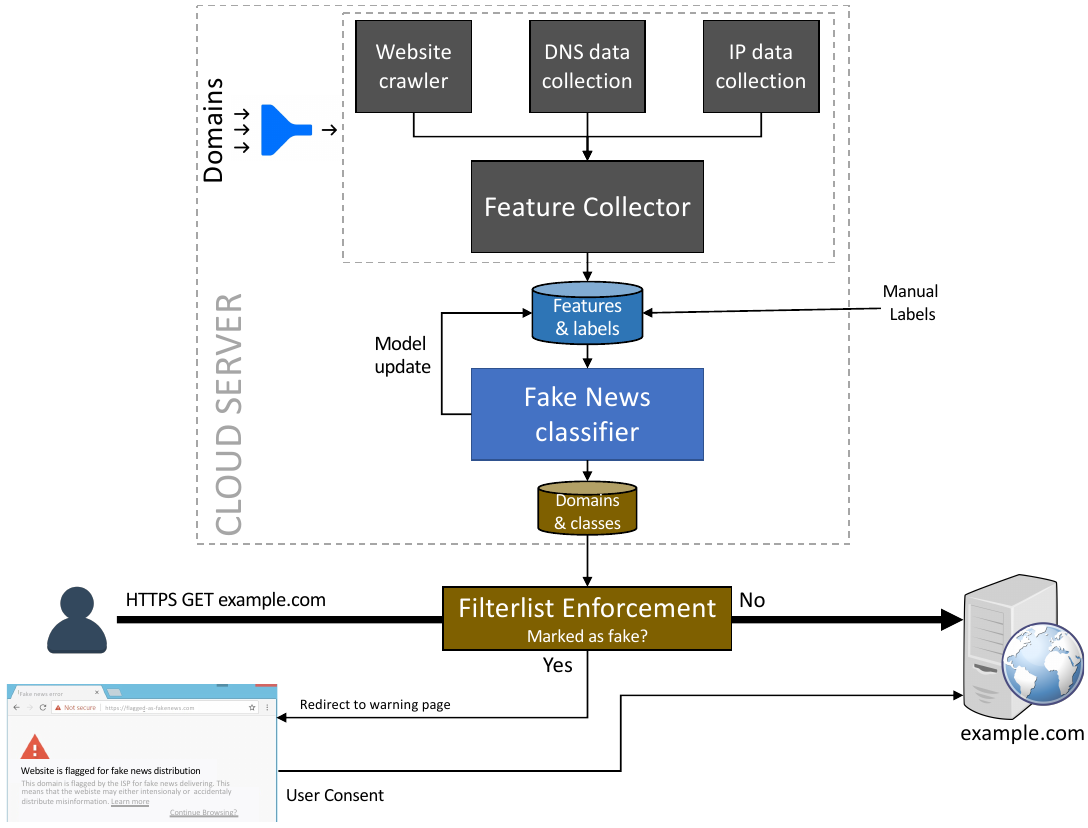}
    \caption{High level overview of \toolname as an ISP service, where the filterlist is applied on the ISP network traffic directly, and users are alerted of fake news website browsing via a web redirection to an information website.}
    \label{fig:design2}
    \end{minipage}\vspace{-0.5cm}
\end{figure}

\point{A) User side filtering}
\toolname can be deployed as a centralized service that is in charge of building a model to  infer when a website is a fake or real news one, and build a filterlist for this purpose.
This list is disseminated to all participating users, and they (their browser) can evaluate whether a visited website is fake news or not.
The server can be maintained either by a private company, where users pay for the service, or a non-profit organization.
As seen in Fig.~\ref{fig:design1}, in this scenario, the \toolname consists of two main components: (1) a cloud server that is responsible for the filterlist generation and maintenance,
and (2) a browser extension that is responsible for retrieving periodically the updated filterlist, and use it to flag the visited website as fake news or not.

For the filterlist generation, the server employs a number of different interconnected modules that include the \emph{Feature Collector}, the \emph{Fake News Classifier}, and the \emph{Filterlist Push Service}.
The \emph{Feature Collector} controls numerous data collecting modules that collect information on websites by:
\begin{enumerate}
\item crawling each website (\eg its DOM structure, SSL characteristics, HTTP headers/redirections and status codes, number of 1st and 3rd party requests to analytics/advertisers, rendering timings, number of included images/fonts/css, number of analytics IDs shared with other affiliated websites, \etc),

\item retrieving data about each website's historical DNS activity (\eg frequency of zone file changes, zone age, number of DNS transactions, historical domain hosts, amount of information provided by the registrant, number of times the domain was parked, \etc),

\item retrieving data about the activity of the website's IP host (\eg frequency of IP changes of host, number of neighboring domains of the same host, geolocation of host's IP, \etc).
\end{enumerate} 

The data collection process of the \emph{Feature Collector} is continuous, since \toolname's plan is to progressively extract features for every site on the web. 
The \emph{Fake News Classifier} is trained to detect fake news sites based solely on features extracted by their structural and network characteristics without attempting any content related analysis.
The classifier gets as input the features extracted, selected and stored by the previous, Feature Collector module.
The results of the classifier form a filterlist which includes the currently detected fake news sites across the web.
This filterlist can then be updated periodically (or upon release of crucial changes), and the \emph{Filterlist Push Service} is responsible for compressing and delivering the deltas (\ie changes from previously deployed version) to the users' browser extensions.

To keep the classifier's accuracy high, the \emph{Model re-trainer} component collects labels (\ie fake/not fake content) manually provided by users who contribute them via their browser extension.
For a new label to be considered as reliable for a given website, it needs to be similarly reported by a certain number of different users.
This helps \toolname to avoid users that maliciously or accidentally report wrong labels.

\noindent
This user-side filtering approach provides some benefits:
\begin{itemize}
\item Preservation of User Privacy: The websites a user is browsing through can expose important information about their interests/preferences, which in some cases can be sensitive (\ie political/sexual preferences, religious beliefs, health issues).
In \toolname, the websites' data processing and classification take place in a centralized manner on the cloud, thus composing a global filterlist with the outcome.
However, the enforcement of this list takes place locally, on each user’s web browser, to avoid leaking to the server the websites visited by the user.

\item Negligible Response Time: Delivering the filterlist to the users' browsers for user side fake news filtering guarantees a practically negligible response time. The same strategy has been proven to provide the best possible user experience in other works on different application domains such as in CRL~\cite{cooper2008internet}, CCSP~\cite{8057065,10.1145/3392096} for certificate revocation, or in~\cite{10.1145/3366423.3380239} for ad-blocking.
\end{itemize}

\noindent
Unfortunately, this implementation has also some downsides:
\begin{itemize} 
    \item The filterlist is an ever-increasing data structure that needs to be updated by the server and efficiently delivered to the user browsers without consuming high network and data bandwidth, otherwise the users may opt for fewer updates.
    \item The filterlist is updated on user browser extensions in waves, leading to the application of potentially deprecated lists for a period of time.
\end{itemize}

\point{B) ISP-side filtering}
Similar to current web filters of ISPs~\cite{webfilters} for filtering websites with hate speech, violence, pornography, phishing, malware, \etc, an alternative design of \toolname is to be deployed as a service provided by their last mile ISP, as shown in Fig.~\ref{fig:design2}.
In this case, the ISP is responsible not only for maintaining the cloud service that generates the periodic filterlist, but also for the automatic enforcement of such a list on the network traffic of its users, at real-time. 
This way, the ISP (who already has access to the browsing history of the user) can monitor and automatically block access (\eg as in the case of phishing/terrorist sites, \etc) or redirect the user's browser to a page that warns them about the flagged-as-fake news site they are about to browse.
The users then can choose to continue with their visit, or to move away from the site. 
Crucially, the \toolname ISP-side approach works on HTTPS traffic, since the approach only takes into account the top-level domain visited by the user, and not the exact URL and the article content.
Overall, this deployment scenario allows \toolname to: 
\begin{itemize}
    \item be immediately applicable as it is deployed centrally by the ISP, and requires zero effort from the users, removing the necessity to download, install and keep up-to-date a user browser extension.
    \item protect more users at once. Similar to the case of phishing and malware sites, ISP-side blocking may be critical for protecting (unaware) customers and their infrastructure or well-being from harmful content.
\end{itemize}

\subsection{Fake News Classifier}
The classification problem at hand can be assumed binary: \ie if a news website can be considered as serving fake or real news content.

\noindent \textbf{Feature Collection and Preprocessing.}
In our study, the features we develop and use for the classification task depart from the state-of-art, and instead are focused on the network, traffic and structural characteristics of each website.
In particular, we analyze features reflecting the various types of data collected, as described earlier. The data collected for these features need to be preprocessed before they can be used to train a machine learning classifier.
This preprocessing involves several steps, including: 
(1) the summarization of HTTP statuses to top levels (\eg 200, 201, 202, \etc, into 2XX),
(2) the imputation of missing data values, where appropriate, and
(3) the normalization of scores within each batch of data used at training.

\noindent \textbf{Feature Selection.}
A total of \textbf{\numOfFeatures} features are computed per website, constructed by the data collected from the different sources.
But not all features have the same importance in this classification task; a typical feature selection process takes place before the ML classifier is trained, to keep the features that contribute the most in solving the specific binary classification problem.
This step is crucial, as it can remove features that not only do not correlate with the class labels at hand, but instead may be adding noise to the whole classification process.
After the feature selection is performed, top \textbf{K} features are selected, based on their contribution to the classification problem.

\noindent \textbf{ML algorithms.}
The fake news classifier can be any standard ML method such as Random Forest, Logistic Regression, Naive Bayes, \etc or more advanced Deep Learning method.
All these methods are batch-based processing ML classifiers trained on a batch of data already collected and pre-processed at the server side, and can be considered ``frozen'', (\ie they cannot change through time).

\begin{figure*}[t]
    \centering
    \begin{minipage}[t]{0.2\textwidth}
        \centering
        \includegraphics[width=1.09\textwidth]{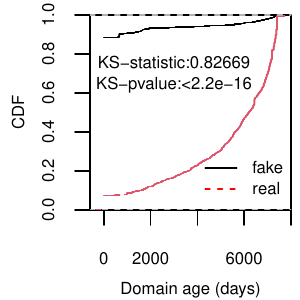}
        \caption{Number of days a given domain has been in existence.}
        \label{fig:domain-age-days}
    \end{minipage}
    \hfill
    \begin{minipage}[t]{0.2\textwidth}
        \centering
        \includegraphics[width=1.09\textwidth]{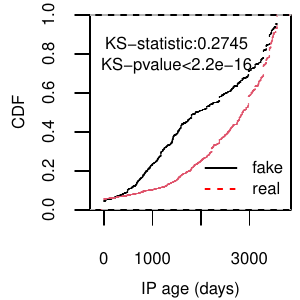}
        \caption{Number of days a given IP has been associated with a specific domain.}
        \label{fig:IP-age-days}
    \end{minipage}
    \hfill
    \begin{minipage}[t]{0.2\textwidth}
        \centering
        \includegraphics[width=1.09\textwidth]{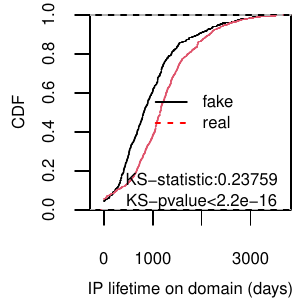}
        \caption{Maximum number of days a given domain had the same IP.}
        \label{fig:IP-change-after-max}
    \end{minipage}
    \hfill
    \begin{minipage}[t]{0.2\textwidth}
        \centering
        \includegraphics[width=1.09\textwidth]{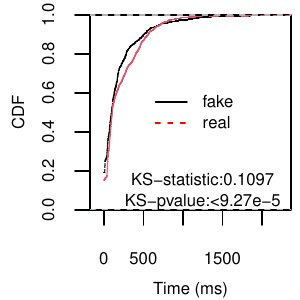}
        \caption{Time duration of user agent to server connection to retrieve page resources.}
        \label{fig:connectDiff}
      \end{minipage}\vspace{-0.4cm}
   \end{figure*} 

\section{Data Collection and Crawler}
\label{sec:dataset}

To obtain the necessary ground truth for our classifier, we utilize the most popular list of manually curated websites of Media Bias Factcheck (MBFC)~\cite{mbfc}. Fact checks of MBFC are carried out by independent reviewers who are associated with the International Fact-Checking Network (IFCN)~\cite{mbfc_methodology} and follow the International Fact-Checking Network Fact-checkers’ Code of Principles, which was developed by the Poynter Institute~\cite{poynter}.

As presented in Table~\ref{tab:dataset}, we retrieve a set of \totalSites that includes \realNewsSet and \fakeNewsSet sites manually verified as real and fake news, respectively.
By developing a puppeteer~\cite{puppeteer} crawler, we visit each one of these sites and extract from its landing page the website DOM tree, the HTTP traffic induced by our crawler's visit, the cookie jar of the website, the memory usage patterns of the JS engine (\eg total and used JS heapsize), the resources the website uses (\eg images, css, js) and the size for each in Bytes, any HTTP headers, the number of included frames and JSEventListeners, the number of $1\times1$ pixels (\ie web beacons) used, and the website's rendering timings.

Next, we retrieve historical metadata about the domain registration, DNS record information and propagation, accessibility from different geolocations via ping, the IP hosting server, other websites using the same Analytics IDs, other websites hosted by the same server and other data, from third party services such as Hosterstats~\cite{hosterstats}, Whoisology~\cite{whoisology}, Spyonweb~\cite{spyonweb} and Viewdns.info~\cite{viewdns}).
Finally, we store all collected data in a PostgreSQL database for further analysis.

As a next step, we process the raw collected data to:
\begin{enumerate}
\item Classify the HTTP traffic to Advertising, Analytics, Social, 3rd-party widgets by using the Disconnect list~\cite{disconnect};
\item Extract the HTTP redirection patterns and the rest of the responded status codes;
\item Extract the frequency of IP changes of the hosting server;
\item Identify the age of the domain, the number of times it was parked and re-registered;
\item Differentiate the first and third party cookies;
\item Study the distribution of a page's different DOM elements (number of divs, buttons, paragraphs, links \etc).
\end{enumerate}
Consequently, we create a vector of \numOfFeatures features for every real and fake news website in our dataset.

\begin{table}[t]
    \centering
    \caption{Summary of the collected dataset.}\vspace{-0.2cm}
    \footnotesize
    \begin{tabular}{lr}
        \toprule
        \bf Type & \bf Volume \\
        \midrule
        Total number of news sites crawled &  \totalSites \\
        News sites labeled as Real & \realNewsSet \\ 
        News sites labeled as Fake &  \fakeNewsSet\\
        Total number of features created &  \numOfFeatures \\
        \bottomrule
    \end{tabular}
    \label{tab:dataset}
    \vspace{-0.3cm}
\end{table}

\section{Data Analysis}
\label{sec:analysis}

In this section, we outline the process followed to select important features that are useful for the training of the ML classifier that will detect fake from real news websites. Then, we perform an in-depth analysis of these top features and how fake news websites differ from real news websites in each of these dimensions.

\subsection{Feature Engineering \& Selection}
\label{sec:feature-eng}
We have collected a large number of features per website, each describing or measuring the specific website in a different way. 
However, several of these features can be highly correlated if they measure closely-related concepts (\eg time/connection-related or size-related aspects).
Such highly correlated features may express the same variability with respect to the labels at hand (fake or real) and are providing the ML classifier with the same information.
Therefore, they can be considered redundant and can be removed, as they do not help the ML classifier, and in some cases, may even add noise to the modeling effort.
Several methods exist for selecting such top contributing features for an ML modeling effort.
In our case, we applied the Recursive Feature Elimination ($RFE$) algorithm available in sklearn library\footnote{\url{https://scikit-learn.org/stable/modules/generated/sklearn.feature_selection.RFE.html}}, with a Logistic Regressor as Fitted Estimator.

As explained earlier, 
we extracted \numOfFeatures features per website, from our collected data across the different sources.
We ran the $RFE$ algorithm to select the top $K$, with $K~\in~\{5, 10, \dots, \numOfFeatures\}$.
We identified that a subset of $K=35$ features provide the best trade-off between ML performance and size of feature set to be processed.
The top features selected can be found in Table~\ref{tab:top-features} and  we see that there are important features from all data source types.
That means, our classification system would need these different data crawls, for the websites to be effectively classified.

\begin{table}[t]
    \centering
    \caption{Top 35 features selected with $RFE$ algorithm.}\vspace{-0.2cm}
    \footnotesize
    \begin{tabular}{p{1.7cm}p{6.5cm}}
    \toprule
    \bf Data Type               &  \bf  Feature \\
    \midrule
    DNS-related activity    & domain\_birth, domain\_age\_days  domainLookupStart, domainLookupEnd  \\ \hline
    IP-related activity     & IP\_change\_after\_max, IP\_age\_days, total\_coownedSites, numOfsites\_coowned\_analytics\\ \hline
    DOM-related details     & domLoading, domContentLoadedEventStart, domContentLoadedEventEnd, domComplete, domInteractive \\ \hline
    HTTP-related details   & connectStart, connectEnd, responseStart,  responseEnd, requestStart, fetchStart, secureConnectionStart, loadEventEnd, loadEventStart \\ \hline
    HTML-related details    & LayoutObjects, Nodes, JSHeapUsedSize,  JSHeapTotalSize, FirstMeaningfulPaint,  HTML\_classes, \{page, image, css, text, js, audio, video\}\_size\\
    \bottomrule
    \end{tabular}\vspace{-0.2cm}
    \label{tab:top-features}
\end{table}

\begin{figure*}[t]
    \centering
    \begin{minipage}[t]{0.2\textwidth}
        \centering
        \includegraphics[width=1.09\textwidth]{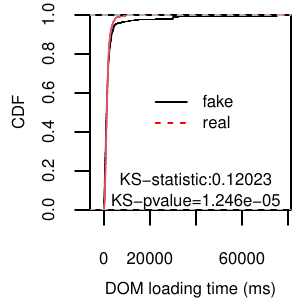}
        \caption{Time for parser to load DOM.}
        \label{fig:domLoading}
    \end{minipage}
    \hfill
    \begin{minipage}[t]{0.2\textwidth}
        \centering
        \includegraphics[width=1.09\textwidth]{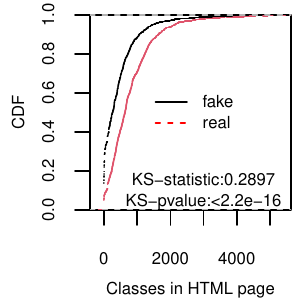}
        \caption{Number of HTML classes in webpage.}
        \label{fig:HTML-classes}
    \end{minipage}
    \hfill
    \begin{minipage}[t]{0.2\textwidth}
        \centering
        \includegraphics[width=1.09\textwidth]{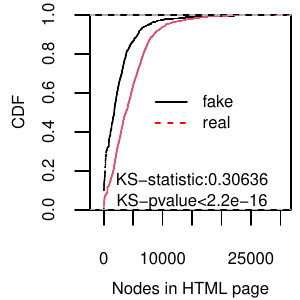}
        \caption{Number of HTML nodes in webpage.}
        \label{fig:Nodes}
    \end{minipage}
    \hfill
    \begin{minipage}[t]{0.2\textwidth}
        \centering
        \includegraphics[width=1.09\textwidth]{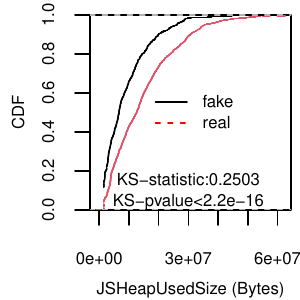}
        \caption{Size of JS heap used by the webpage.}
        \label{fig:JSHeapUsedSize}
    \end{minipage}\vspace{-0.5cm}
\end{figure*}
\begin{figure*}   
    \begin{minipage}[t]{0.2\textwidth}
        \centering
        \includegraphics[width=1.09\textwidth]{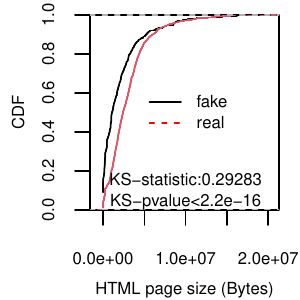}
        \caption{Total webpage size.}
        \label{fig:page-size}
    \end{minipage}
    \hfill
    \begin{minipage}[t]{0.2\textwidth}
        \centering
        \includegraphics[width=1.09\textwidth]{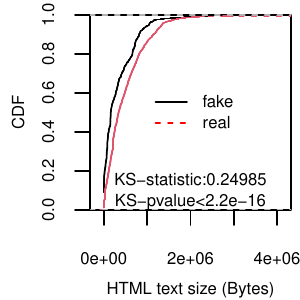}
        \caption{Total page text size.}
        \label{fig:text-size}
    \end{minipage}
    \hfill
    \begin{minipage}[t]{0.2\textwidth}
        \centering
        \includegraphics[width=1.09\textwidth]{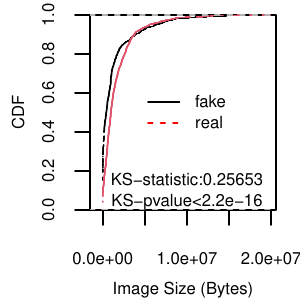}
        \caption{Total images size.}
        \label{fig:image-size}
    \end{minipage}
    \hfill
    \begin{minipage}[t]{0.2\textwidth}
        \centering
        \includegraphics[width=1.09\textwidth]{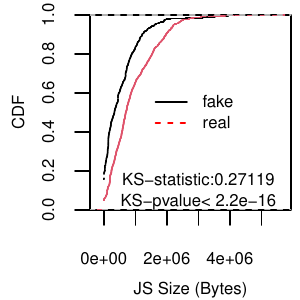}
        \caption{Total JS code size.}
        \label{fig:js-size}
    \end{minipage}\vspace{-0.5cm}
\end{figure*}

\subsection{Top Feature Analysis}
\label{sec:feature-analysis}
We study the summary statistics of these top 35 features and discuss differences between the two labels (fake or real news websites).
We also perform a statistical analysis on each of the selected features to understand if they are statistically different, using non-parametric Kolmogorov-Smirnov tests. 
We report the results for each test (test-statistic and associated p-value) in each figure.
Due to space, here we only analyze 12 out of the 35 top features.

\point{DNS-related features}
Fig.~\ref{fig:domain-age-days} shows the distribution of the number of days that a domain from the set of fake or real news websites has been in existence on the Web.
We immediately notice that the domains of the fake news websites have a much smaller median lifetime than real news websites (0 days vs. 6197 days).
In fact, the great majority of fake news (almost 90\%) have close to zero days of their domain being alive.
Interestingly, though not shown here due to space, when we study the number of websites co-owned by the same entity, we find that the two types of news websites do not differ in a statistically significant manner.

\point{IP-related features}
Fig.~\ref{fig:IP-age-days} shows the distribution of the number of days that a given IP used by a website has been associated with the specific website and domain.
Again, we notice a significant difference between fake and real news websites, with median 815.5 days vs. 1148.0 days, respectively.
Also, in Fig.~\ref{fig:IP-change-after-max}, we show the maximum lifetime of an IP while being associated with a specific domain.
Similarly with earlier results, we find that IPs associated with real news websites have a higher median such lifetime (2857 days) vs. fake news websites (1890 days). 
    
\point{HTTP-related features}
As a representative of the network activity and delays expected while browsing a fake or real news website, in Fig.~\ref{fig:connectDiff}, we study a compound metric that reflects the time difference between {\tt connectEnd} and {\tt connectStart}. 
In other words, this metric quantifies how long it takes for the user agent to retrieve resources from a specific website domain server.
We find that, though the distributions are very similar (\ie medians are similar with 98ms vs. 105.5ms for fake vs. real, respectively), they are still statistically different (p-value=$9.27e^{-5}$).

\point{DOM-related features}
Similarly, we pick the \emph{domLoading} feature to study DOM-related characteristics of websites. 
Fig.~\ref{fig:domLoading} shows the time taken for the DOM tree of a website to be loaded.
Again, both types of websites appear to have similar distribution, especially in median numbers (1179ms vs. 1306ms for fake vs. real, respectively).
However, they are still statistically different (p-value=$1.246e^{-5}$), due to the $5\%$ tail of the fake news websites, which are demonstrating longer load times for the DOM tree than real news websites.

\point{HTML-related features}
Lastly, we take a look at features reflecting HTML properties of the websites, and in particular, characteristics such as how many classes (Fig.~\ref{fig:HTML-classes}) and nodes (Fig.~\ref{fig:Nodes}) are in the HTML of the page, Javascript heap size used (Fig.~\ref{fig:JSHeapUsedSize}), as well as overall size of the webpage (Fig.~\ref{fig:page-size}), text (Fig.~\ref{fig:text-size}), image (Fig.~\ref{fig:image-size}) and Javascript code (Fig.~\ref{fig:js-size}) included in the page.
We notice that in all these metrics, real news websites have higher median than fake news websites, and the two distributions are always statistically different with p-value $<2.2e^{-16}$.
In particular, real news websites have more than twice as high median number of classes in their HTML page than fake news (638 vs. 262 classes), as well as HTML nodes (3760 vs. 1778 nodes).
Real news websites also use almost double the Javascript heap than fake news (medians: 12.03MB vs. 6.37MB).
Interestingly, real news websites have more than double the median overall page size compared to fake news (2.31MB vs. 1.03MB).
This is similarly reflected in the overall text included (332.42KB vs. 154.99KB), images included (926KB vs. 402KB), and Javascript code included (680.63KB vs. 327.8KB).

We find some extreme cases of fake/real news websites that have a very large number of classes and/or nodes in their HTML page:
\begin{itemize}
    \item \url{newnation.org}: Has the highest number of nodes (29419) for a fake news site in our list, but only a small number of classes (119).
    \item \url{strategypage.com}: Has the highest number of nodes (207914) for a real news site in our list, and a large number of classes (7171).
    \item \url{uft.org}: Has the highest number of classes for a real news site (8058) in our list, and a large number of nodes (29052), close to the largest in fake news (29419).
    \item \url{loser.com}: Records the highest number of classes for a fake news website (5216) in our list, but also a very large number of nodes (26385) close to the largest in fake news (29419).
\end{itemize}

\section{ML-based Fake News Detection}
\label{sec:ml-classifier}

In this Section, we describe our effort to train and test various ML classifiers on the collected data, and demonstrate the efficacy of our method to infer fake or real news websites at almost real-time using run-time features.

\subsection{Asynchronous vs. Real-time ML Classification}

For the training and testing, we performed a split of the data ($35$ selected features for the $1822$ websites) into $80\%$ for training and validation (by applying 5-fold cross-validation method), and $20\%$ for testing on unseen data.
Example classifiers for this process can be either classic ML methods such as Random Forest or Logistic Regression, as well as Neural Networks ($NN$).
Next, we outline experiments we did with both types of classifiers.

\point{NN-based Classifier}
A simple NN classifier for this problem can have the following layered architecture:
\begin{enumerate}
    \item[L1:] Feature/Input layer
    \item[L2:] Batch Normalization layer
    \item[L3:] Dense Layer (\emph{relu} activation function) with X neurons
    \item[L4:] Dropout layer (rate=0.1)
    \item[L5:] Classification/Output layer
\end{enumerate}
X is varied in the range of 8, 16, 32, \dots, 128.


We implemented the above mentioned $NN$ architecture using the $Tensorflow$ $Keras$ framework, version 2.5.0.
We allowed the network to train for 10 rounds and 50 epochs each.
We measured average accuracy and average loss for each round.
The configuration with optimal results offered an average accuracy of $92.44\%$ and average loss of 0.2436.

\begin{table}[t]
    \centering
    \footnotesize
    \caption{Top: Performance metrics from ML binary classification of news websites: 0: likely to be hosting real news content; 1: likely to be hosting fake news content.
    Bottom: Baseline/State-of-art ML methods for binary classification of fake news articles and websites (metrics reported from each past work.}\vspace{-0.2cm}
    \begin{tabular}
    {p{3.5cm}p{0.36cm}p{0.36cm}p{0.36cm}p{0.36cm}p{0.36cm}p{0.36cm}}
    \toprule
    \bf Method   &   \bf TPRt &  \bf FPRt & \bf  Prec.   &  \bf Recall   &   \bf \;\;F1      &  \bf AUC     \\ \midrule
    Random Forest (RF)      &   0.916   &   0.099   &   0.916       &   0.916   &   0.916   &   0.967   \\
    Logistic Regression &   0.909   &   0.109   &   0.909       &   0.909   &   0.909   &   0.944   \\
    Naive Bayes        &   0.805   &   0.271   &   0.801       &   0.805   &   0.800   &   0.856   \\
    Multi-layer Perceptron\ & & & & & &\\
    Neural Net (NN) (35x20x2)    &   0.912   &   0.107   &   0.912       &   0.912   &   0.912   &   0.936   \\ \midrule
    Linguistic-based SVM~\cite{SAHOO2021106983}  & ---&   ---   &   0.730  &  0.730   &   0.725  &   ---  \\
    Linguistic-based Ensem.~\cite{uddin2020ensenbles} & --- & ---    &  0.953	&  0.940 &  0.945 & --- \\
    User-engage-based RNN~\cite{10.1145/3132847.3132877} & ---	&  ---   &  ---	&   ---     &  0.924 &   ---	\\
    Linguistic-based DNN~\cite{paschalides2019check}    & ---	&  ---   & 0.719 & 0.714  &  0.711 &   ---  \\
    SocialNets-based RF~\cite{10.1145/3373464.3373473} & --- & ---    &  ---	&   ---     &  0.887 & --- \\
    \bottomrule
    \end{tabular}
   \label{tab:classification-results-top35}
\end{table}

\begin{table}[t]
    \footnotesize
    \centering
    \caption{Performance metrics from ML binary classification of news websites, using real-time available features only.}\vspace{-0.2cm}
    \begin{tabular}
    {p{3cm}p{0.6cm}p{0.6cm}p{0.4cm}p{0.4cm}p{0.4cm}p{0.4cm}}
    \toprule
     \bf Method  &  \bf TPRate &  \bf FPRate & \bf Prec. & \bf Recall & \bf \;\;F1  &  \bf AUC\\ \midrule
    Random Forest  &   0.819   &   0.294   &   0.826       &   0.819   &   0.808   &   0.867   \\
    Logistic Regression &   0.747   &   0.378   &   0.741       &   0.747   &   0.732   &   0.747   \\
    Naive Bayes &   0.668   &   0.478   &   0.648       &   0.668   &   0.647   &   0.656   \\
    Neural Net     (27x128x2)   &   0.745   &   0.376   &   0.738       &   0.745   &   0.731   &   0.769   \\ 
    \bottomrule
    \end{tabular}
    \label{tab:classification-results-top-real-time}\vspace{-0.3cm}
\end{table}

\point{Classic ML Classifiers}
For completeness, we also experimented with classic ML classifiers and provided a comparison of their performance in Table~\ref{tab:classification-results-top35}, on different performance metrics, as weighted averages across the two classes.
Looking at the performance of these methods, we note that in comparison with the more advanced NN-based classifiers, even classic ML classifiers like Random Forest or Logistic Regression can perform very well in classifying the fake from real news websites, when good features are available.
In fact, Random Forest demonstrates very high True Positive (TR) and very low False Positive (FP) rates, very high Precision, Recall and F1 scores, and very good Area Under the Receiver Operating Curve (AUC), which compares remarkably well with the NN classifier, and with even better AUC than NN.

\point{Comparison with Baseline Methods}
In the bottom half of Table~\ref{tab:classification-results-top35}, we also report performance of various state-of-art methods for metrics shown in the respective studies (note: if a method was evaluated on different datasets, we compute average performance per metric).
Our methodology fares comparatively well against these past works that rely on social network, linguistic or user engagement features, but not network traffic features, to detect sites distributing fake news.
Overall, the results from the classical and NN ML methods, and how they compare with past studies, demonstrate that our methodology of collecting and selecting such data features can be of crucial value in classifying websites serving fake from real news, without the need to analyze the content of the articles themselves.

\point{Real-time Detection}
Next, we attempt a classification task which uses only the data per website that is available by the system at run-time, in order to perform real-time detection of fake or real news.
This means the system (and to an extent the ML classifier) cannot use data sources that require previous crawling and processing, nor data sources which need to be crawled at run-time.
Instead, the system would use this real-time ML classifier to assess the label of the website based on the data that are available during the access of the website by the end user.
Of the top 35 attributes found to be most helpful in the binary ML task, 27 are available at run-time.
That is, we exclude the 8 features related to IP and DNS activity, and repeat the ML classification task with the rest of top features.

Table~\ref{tab:classification-results-top-real-time} demonstrates the results for some basic ML classifiers, as well as the NN-based classifier.
We find that a Random Forest classifier can maintain a competitive performance with more than 81\% Precision and Recall, in detecting fake from real news websites, and in fact performs better than some simple NN-based ML classifier.
In particular, these results show that our methodology, data features considered and our overall pipeline proposed can detect fake news websites even at run-time with good accuracy (almost 82\%).
Indeed, while the end users are browsing the Web, there is no time to fetch extra information about each browsed website, but instead, the ML classifier has to rely on the existing data features extracted from the website browsed, \ie from feature categories related to the DOM and webpage structure, as well as HTTP traffic-related activity.
Of course, extra information can be collected in the background per new website seen by the system, in order to enrich the ground truth dataset and allow the ML classifier to perform even better, \ie reach the performance scores (see previous subsection) on the asynchronous ML classification.

\subsection{Validation on Newly-Flagged Websites}
Finally, we perform a validation study on 100 websites, newly-flagged by MBFC~\cite{mbfc} (Nov. 2022) as serving fake or real news.
Following the same process as earlier, we crawl these websites and their respective sources of metadata, perform feature cleaning and selection and focus on the top 35 as selected earlier, for ML classifier training and testing.
Using the Random Forest classifier trained earlier, 
we attempt to classify these 100 websites.
Since the classification is binary, we rank websites based on the probability to be flagged as fake or real, and focus on the top ranked ones.

The classifier flagged 19/100 websites as fake news with probability above 0.5.
We manually confirm that 14/19 (74\%) are indeed serving such news.
The classifier also detects an interesting cluster of fake news websites that are similarly styled and named (*catholictribune.com), where * denotes the name of different US states, and reflects the type of clusters detected in a past study~\cite{10.1145/3447535.3462510}.
From the remaining 81 websites, the classifier flags 24 as real news with probability above 0.9.
We confirm that 22/24 (92\%) are indeed serving real news.
Also, 7/81 websites are flagged with low probability to be serving real news (prob$<$0.6).
Indeed, we confirm that 6/7 of them serve fake news, raising to 77\% (i.e., (6+14)/(19+7)) the confirmed fake news detection.

Overall, these results provide confidence that the classifier we built is useful and can detect with very good accuracy websites that serve real news, as well as successfully flag websites that potentially serve fake news, and need to be investigated further by experts.
\section{Prototype Implementation}
\label{sec:implementation}

To assess the effectiveness and feasibility of our approach, we developed a first prototype of \toolname.
Our system, as earlier illustrated in Fig.~\ref{fig:design1}, includes two main components: (i) the server that crawls and classifies websites (described in Section~\ref{sec:design}), and (ii) \extension: a browser extension that warns the users about the credibility of the website they are about to view.
Both components are publicly released for extensibility and research repeatability. 

\subsection{\toolname Server}
The server consists of 4 main components:
\begin{enumerate}
    \item The crawler component utilizes Puppeteer~\cite{puppeteer} to crawl large portions of websites, while collecting their metadata (\ie DNS and IP information) as described in Section~\ref{sec:dataset}.
    \item Sqlite3 database that stores all collected data per website
    \item ML module (written in Python) that extracts features out of the collected information and classifies each crawled website. Additionally, the ML component retrains the model periodically based on user label contributions.
    \item The server's endpoint is responsible for (a) creating and distributing a Fake News filterlist and (b) their deltas, while it also (c) receives the label contributions of authorized users.
\end{enumerate}

\begin{figure}[t]
    \begin{minipage}[t]{0.47\textwidth}
        \centering
        \includegraphics[width=.7\textwidth]{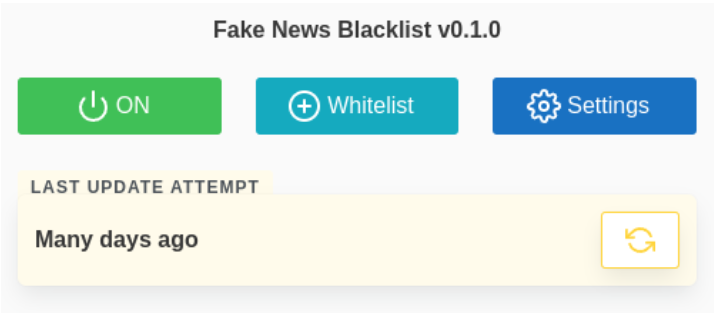}
        \caption{\extension users are presented with a menu to turn on/off the tool, view/change its settings or whitelist arbitrary domains. They can also manually update their local filterlist.}
        \label{fig:menu}\vspace{-0.0cm}
    \end{minipage}
    \hfill
    \begin{minipage}[t]{0.47\textwidth}
    \centering
        \includegraphics[width=.7\textwidth]{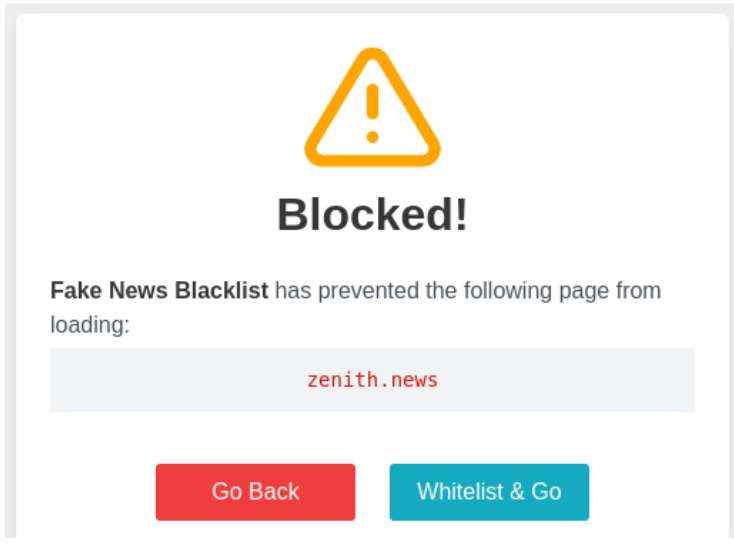}
        \caption{If the visited domain is blacklisted, the users are presented with a warning page and the choice to go back, or whitelist the visited domain.}\vspace{-0.0cm}
        \label{fig:blocked}
    \end{minipage}
\end{figure}

\point{The Fake News filterlist}
This filterlist includes (i) websites that have been checked and whitelisted (flagged as credible news sites) and (ii) websites that have been checked and blacklisted (flagged as fake news sites).

\subsection{\extension extension}
This component is responsible for enforcing the server's filterlists on the users' browsers. Specifically, it is a Chromium-based browser extension (Fig.~\ref{fig:menu} illustrates the provided menu) that blocks fake news domains for end users. 
The filterlist is fetched (in the form of JSON) from the server either entirely (\ie bootstrapping phase), or in the form of deltas (\ie the changes that took place since the last time it was fetched): as soon as the users open their browsers, the extension checks for possible updates it may have missed and fetches the deltas from a specific checkpoint onwards.

The filterlist is stored locally in the browser's LocalStorage and is queried every time a user is visiting a new domain. The filterlist is sorted and binary search is used to find a domain in the list, so the querying complexity is $O(logn)$. Whenever the user visits a new domain, the extension checks if the domain is white- or black- listed. If the latter, the user is presented with a warning page (as shown in Fig.~\ref{fig:blocked}) 
and has the choice to go back, or whitelist the visited domain. Our browser extension is written in JavaScript and makes use of the Bulma CSS~\cite{BulmaCSS} library for styling, Umbrella JS~\cite{umbrellaJS} for DOM Manipulation.

\begin{figure}
    \centering
    \includegraphics[width=0.3\textwidth]{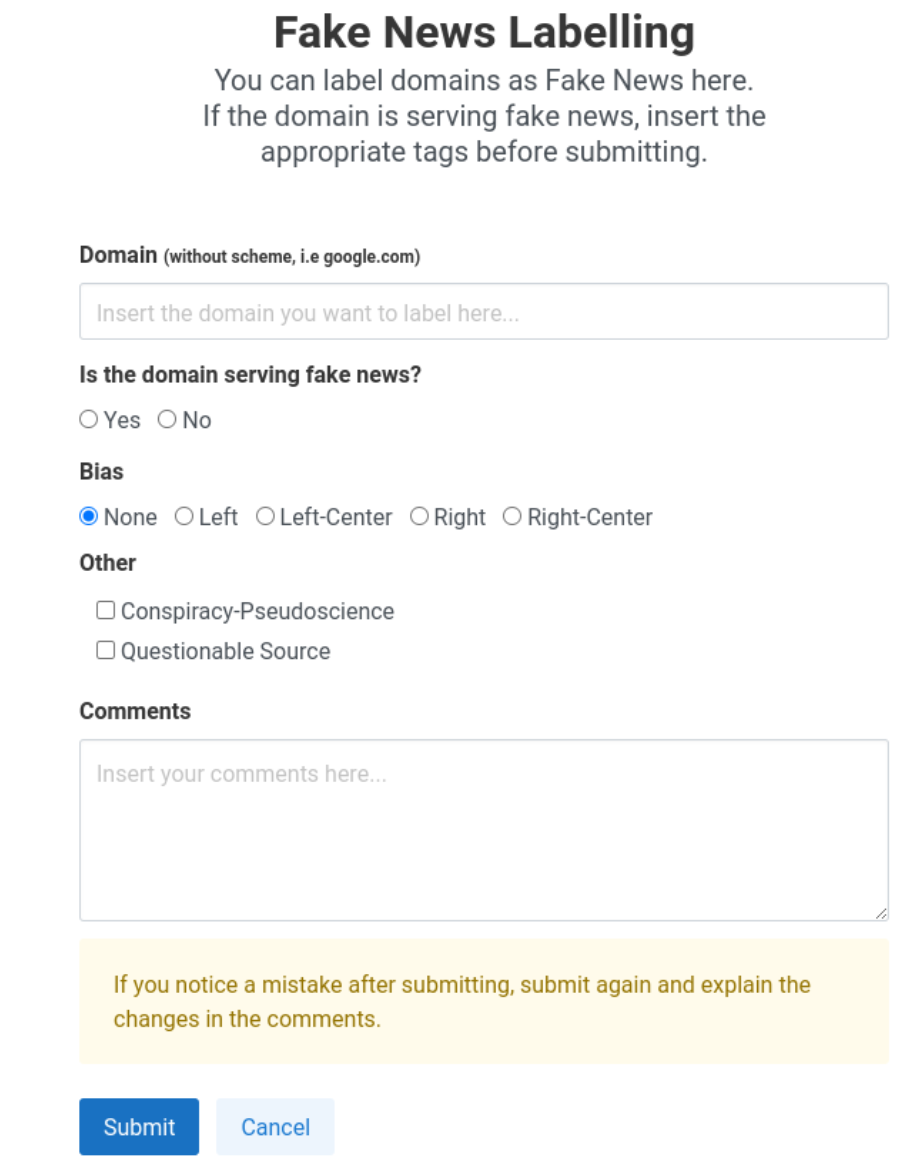}
    \caption{\extension provides a form to authorized users, where they can blacklist, report or whitelist a website.}\vspace{-0.1cm}
    \label{fig:contributions}
\end{figure}

\point{Label Contributions}
In \extension, authorized users can contribute with their labels when they think a website should be flagged as fake news or credible.
Thus, there is a specific functionality provided (as shown in Fig.~\ref{fig:contributions}), 
where users can whitelist a flagged website or blacklist and report a website they consider as fake news. The results of these actions are not only applied locally in their local filterlist, but in the case of \emph{super-users} (\ie that have been manually curated, authorized and considered trusted) they are also pushed to the remote server together with a JSON Web Token (JWT authentication). The remote server aggregates this information and after basic processing to verify the report validity, it can use them as feedback (\ie manual labels) to retrain and improve its ML model\footnote{In this prototype, we perform a basic filtering for possible authorized users that turned malicious and abuse the reporting functionality. A more sophisticated analysis and verification was beyond the scope of this prototype.}.

\section{System Evaluation}
\label{sec:evaluation}

In this section, we set out to evaluate the performance of \toolname' client-side component in an attempt to assess its impact on the user experience.
Hence, we deploy \toolname in Google Chrome browser on a device running Void Linux equipped with an Intel Core i5-4440 CPU 
and 8GB DDR3 RAM over a 50 Mbps Internet connection.

\begin{figure}[t]
    \centering
    \begin{minipage}[t]{0.235\textwidth}
        \centering
        \includegraphics[width=1.11\textwidth]{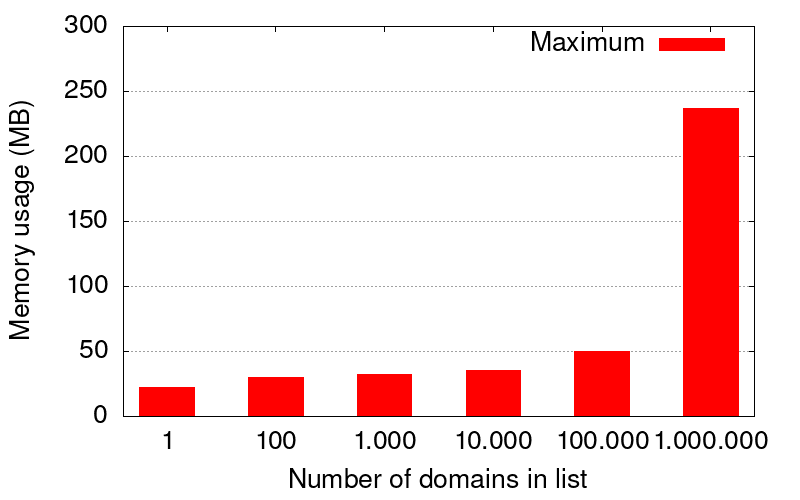}\vspace{-0.1cm}
        \caption{Max additional memory used by \extension per new website visit. For a normal size of 10K-100K filterlist, the memory used is negligible ($<$50MBytes).}
        \label{fig:memory}
    \end{minipage}
    \hfill
    \begin{minipage}[t]{0.235\textwidth}
        \centering
        \includegraphics[width=1.12\textwidth]{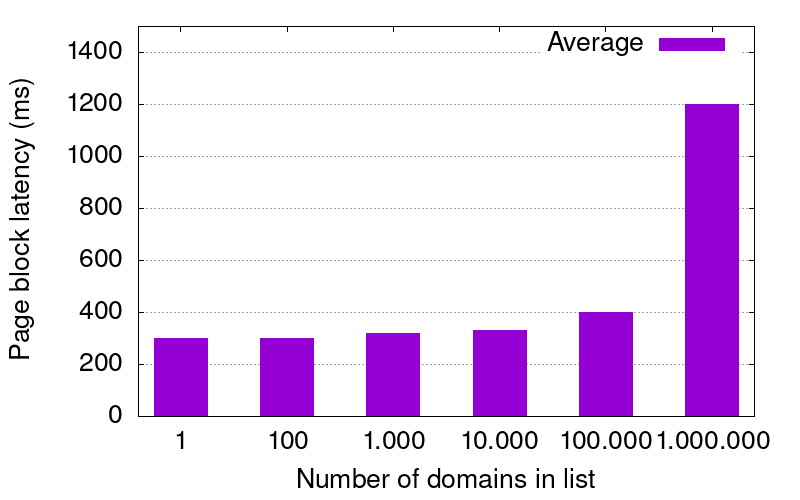}\vspace{-0.1cm}
        \caption{Average end-to-end execution time of \extension (time between the user clicking on a domain flagged as fake news and the warning page rendered).}
        \label{fig:latency}
    \end{minipage}\vspace{-0.3cm}
\end{figure}

First, we measure the additional memory our extension uses for the different size of filterlist. In Fig.~\ref{fig:memory}, we add {\tt google.com} in the filterlist before visiting its landing page (\ie a basic page without dynamic content) and using Chrome's built-in Task Manager, we measure the maximum memory the extension uses for the different filterlist sizes.
We repeat our visits 100 times and plot the averages of this maximum memory consumption.
Before each and every visit, the following actions are taken to ensure a clean slate:
(i) the extension's local storage was reset, (ii) an update was triggered to fetch the list.
(iii) the browser was restarted to clear the cache.
As we see from the figure, for a normal size of filterlist 10,000 - 100,000 websites (the most popular ad-blocking filterlist --- Easylist --- contains 70,000 rules~\cite{easylistSize}), the memory used is less than 50 MBytes, which is practically negligible for today's consumer-targeting systems. 

Then, we measure the extension's end-to-end execution time: from the moment the user clicks on a domain till the warning page (Fig.~\ref{fig:blocked})) 
is rendered in case it is flagged as fake news. We visit again the simple landing page of {\tt google.com} to ensure our measurements are not affected by dynamic content, and by using the network tab in chrome's developer tools (with cache disabled and persistent logging enabled), we measure the time it takes for the extension to query the filterlist and render the warning page. We repeat our experiment 100 times and in Fig.~\ref{fig:latency} we plot the averages. 
It is important to note at this point that our extension works asynchronously and does not impede the loading operation of the webpage: \ie filterlist querying takes place in parallel to the webpage's loading procedure. In Fig.~\ref{fig:latency}, we see that from the moment the user clicks on a link to navigate to a fake news site, till the warning page is rendered on the user's display, it takes, on average, less than 400 ms, which compares favorably to the average page load time found to be higher than 1100ms in past studies~\cite{averageSiteLatency}.

\point{Overall Performance}
The proof-of-concept implementation results demonstrate the feasibility of the envisioned service: the server, coupled with the client-side extension can operate at practically negligible memory and latency costs for a user.

\section{Conclusion}
\label{sec:conclusion}
In this work, we propose \toolname: the first of its kind holistic \emph{Fake News Detection as-a-Service} system that is completely content-agnostic.
\toolname does not take into account the content of the article, but instead relies on the network and structural characteristics of the website (\eg DNS record changes, domain age, domain expiration and re-registration patterns, DOM characteristics, number of HTTP redirections, page rendering timings, \etc).
We evaluate the feasibility and effectiveness of our approach by using a set of \fakeNewsSet fake and \realNewsSet real news websites.
We tested the ML performance of \toolname under different experimental settings.
The performance results of various ML classifiers showed that \toolname can achieve high Precision, Recall, and F1 scores (all above 91\%) and AUC score (0.967) when asynchronously collected features are used. When features available at run-time are used, ML classification is still possible with accuracy performance up to 82\%.
These results mean our \toolname can be applicable even without using data that require a lot of background collection process and time.
Finally, we developed a first prototype of the server and client components of \toolname and demonstrated its efficacy on real user devices.

\section{Ethical Considerations}
\label{sec:ethics}

The execution of this work and data collection has followed the principles and guidelines of how to perform ethical information research and the use of shared measurement data
%
We kept our website crawling to a minimum to ensure that we do not slow down or deteriorate the performance of any web service or provided web API in any way.
Therefore, we crawl only the landing page of each website and visit it only once.
We do not interact with any component in the website visited, and only passively observe network traffic.
Therefore, we make concerted effort not to perform any type of DoS attack to the visited website.
In accordance to the GDPR and ePrivacy regulations, 
we do not share with any other entity any data collected by our crawler.
Regarding the ads in visited websites, we were cautious not to affect the advertising ecosystem or deplete advertiser budgets, by minimizing the number of crawling per website.

Regarding the potential issue of websites reacting to our proposed service that it collects data for labeling them as ``fake'' or ``real'' news without their consent, we argue that any (news) content publicly available on the Web is already up for scrutiny by existing tools, methodologies, journalists, \etc, and then, websites found to be servicing ``fake'', or ``questionable'' \etc news are labelled as such.
Our proposed service, in either of the two facets, just facilitates this process in a more mainstreamed and effective way.



\section*{Acknowledgements}
The research leading to these results received partial funding from the EU H2020 Research and Innovation programme under grant agreements No 830927 (Concordia), No 871370 (Pimcity), No 871793 (Accordion), and No 101132686 (ATHENA). The authors bear the sole responsibility for the content presented in this paper, and any interpretations or conclusions drawn from it do not reflect the official position of the EU, FORTH, or Telefonica.

\begin{spacing}{0.9}
\bibliographystyle{unsrt}
\bibliography{main}
\end{spacing}
\balance
\end{document}